\documentclass[
 reprint,
superscriptaddress,
 amsmath,amssymb,
 aps,
]{revtex4-2}
\usepackage[version=3]{mhchem} 
\usepackage{graphicx}
\usepackage{dcolumn}
\usepackage{bm}
\usepackage{upgreek}
\usepackage{siunitx}
\usepackage{lipsum} 
\usepackage{physics}
\usepackage{soul}
\begin{document}
\preprint{APS/123-QED}

\title{Stroboscopic X-ray Diffraction Microscopy of Dynamic Strain in Diamond Thin-film Bulk Acoustic Resonators for Quantum Control of Nitrogen Vacancy Centers}

\author{Anthony D’Addario}
\affiliation{Department of Physics, Cornell University, Ithaca, NY, USA}
\author{Johnathan Kuan}
\affiliation{Department of Physics, Cornell University, Ithaca, NY, USA}
\author{Noah F. Opondo}
\affiliation{Elmore Family School of Electrical and Computer Engineering, Purdue University, West Lafayette, IN}
\author{Ozan Erturk}
\affiliation{Elmore Family School of Electrical and Computer Engineering, Purdue University, West Lafayette, IN}
\author{Tao Zhou}
\affiliation{Center for Nanoscale Materials, Argonne National Laboratory, Lemont, IL}
\author{Sunil A. Bhave}
\affiliation{Elmore Family School of Electrical and Computer Engineering, Purdue University, West Lafayette, IN}
\author{Martin V. Holt}
\affiliation{Center for Nanoscale Materials, Argonne National Laboratory, Lemont, IL}
\author{Gregory D. Fuchs}
\email{gdf9@cornell.edu}
\affiliation{School of Applied and Engineering Physics, Cornell University, Ithaca, NY, USA}
\affiliation{Kavli Institute at Cornell for Nanoscale Science, Ithaca, NY}


\begin{abstract}

Bulk-mode acoustic waves in a crystalline material exert lattice strain through the thickness of the sample, which couples to the spin Hamiltonian of defect-based qubits such as the nitrogen-vacancy (NV) center defect in diamond.  This mechanism has been previously harnessed for unconventional quantum spin control, spin decoherence protection, and quantum sensing. Bulk-mode acoustic wave devices are also important in the microelectronics industry as microwave filters. A key challenge in both applications is a lack of appropriate operando microscopy tools for quantifying and visualizing gigahertz-frequency dynamic strain. In this work, we directly image acoustic strain within NV center-coupled diamond thin-film bulk acoustic wave resonators using stroboscopic scanning hard X-ray diffraction microscopy at the Advanced Photon Source. The far-field scattering patterns of the nano-focused X-ray diffraction encode strain information entirely through the illuminated thickness of the resonator. These patterns have a real-space spatial variation that is consistent with the bulk strain’s expected modal distribution and a momentum-space angular variation from which the strain amplitude can be quantitatively deduced. We also perform optical measurements of strain-driven Rabi precession of the NV center spin ensemble, providing an additional quantitative measurement of the strain amplitude. As a result, we directly measure the NV spin-stress coupling parameter $b =2.73(2)$ MHz/GPa by correlating these measurements at the same spatial position and applied microwave power. Our results demonstrate a unique technique for directly imaging AC lattice strain in micromechanical structures and provide a direct measurement of a fundamental constant for the NV center defect spin Hamiltonian.

\end{abstract}

\maketitle

\section{Introduction \label{sec:Intro}}

Engineered control over lattice strain -- static \cite{Ovartchaiyapong_2014, Teissier_2014} and dynamic \cite{Macquarrie_2013} --  enables quantum control of atomic-scale point defects for quantum sensing \cite{Barry_2020} and improved qubit performance such as extending spin coherence times \cite{Macquarrie_decoup_2015, Barfuss_2015}, tuning defect state energy levels \cite{Ovartchaiyapong_2014, Teissier_2014}, and driving coherent spin \cite{Macquarrie_2015, Barfuss_2015} and orbital \cite{Lee_2016, Chen_2018} transitions. Surface and bulk acoustic strain waves in the lattice that are coupled to defects have been demonstrated by micro-electromechanical systems (MEMS) that use piezoelectric transducers \cite{Macquarrie_2013, Macquarrie_2015, Chen_2019, Chen_2020, Maity_2020, Maity_2022, Golter_2016, Dietz_2022, Whiteley_2019, Fu_2008, Barclay_2009}. Bulk acoustic wave (BAW) resonators are particularly useful because of their high operating frequencies, high quality factors, and high thermal conductivity enabling robust power handling \cite{Liu_2020, Wang_2022}. Also, BAW resonators are electronic devices, allowing for convenient integration of an additional microwave (MW) control antenna or other quantum control fields. It is especially interesting to quantify the generated bulk acoustic strain in BAW devices because they are widely prevalent in applications including filters, oscillators, and sensors \cite{Liu_2020, Wang_2022} in addition to coupling with defects.

Previous work has shown that acoustic strain waves in high-overtone bulk acoustic wave resonators (HBARs) enable coupling to the nitrogen-vacancy (NV) center defects in diamond \cite{Barry_2020}. Gigahertz frequency lattice strain drives coherent Rabi oscillations between two ground spin states \cite{Macquarrie_2013, Macquarrie_2015, Chen_2019, Chen_2020} and is directly proportional to the Rabi frequency, providing a quantitative measurement of this strain \cite{Chen_2019}. An alternative approach for examining NV-strain coupling uses microbeam resonator systems. For these, cantilever deflection is directly measured, which is converted to strain using Euler-Bernoulli beam theory \cite{Ovartchaiyapong_2014, Teissier_2014}. These works form the most accurate prior measurements of the NV ground state spin axial and transverse strain sensitivities. More direct measurements of lattice displacement are possible using scanning X-ray diffraction microscopy (SXDM), where Bragg diffraction provides a direct measurement of the displacement of atoms in the lattice \cite{Holt_2013}. Experiments at the Advanced Photon Source (APS) using SXDM have studied surface acoustic waves (SAWs) on a SiC lattice with correlated photoluminescence of an ensemble of defects \cite{Whiteley_2019}. Meanwhile, measuring BAW strain is more challenging because the lattice is both compressively and tensilely strained through the thickness. Direct measurements will necessarily sum over both forms of strain, complicating the measurement of the lattice displacement through the bulk. However, a direct measurement is important because it sheds light on the BAW device's behaviors, limitations, and pathways for fabrication improvements, and because it provides the most direct comparison yet between dynamic lattice strain and the corresponding NV center response.

In this work we use time-resolved SXDM (tr-SXDM) at the Hard X-ray Nanoprobe Beamline operated by the Center for Nanoscale Materials (CNM) at the APS to measure gigahertz frequency strains generated in diamond thin-film bulk acoustic resonators (FBARs). We measure the bulk acoustic strain made possible by the penetrating depth of the X-rays, allowing for diffraction summed over the entire depth of atomic displacements \cite{Holt_2013}. The monochromatic nature of the X-ray beam allows us to isolate the response of diamond, ignoring other materials that comprise the FBAR device. We measure strain stroboscopically, allowing for repeatable measurements of the strain wave with twenty picosecond resolution by driving the resonator referenced to the X-ray bunches. Also, we perform confocal microscopy of diamond NV centers to access the strain using an independent method. We measure Rabi oscillations driven by the strain wave at the same drive power and frequency used at the APS. With these two methods of analysis, we make a direct measurement of the spin-stress parameter for NV centers, $b=2.73(2)$ MHz/GPa, that relates strain and Rabi frequency. Our measurement agrees well with previous results, and Table \ref{tab:results} contains the results of our direct measurement in addition to the other three measurements of this parameter.

\section{Device \label{sec:Device}}

\begin{figure}
\includegraphics[scale=1]{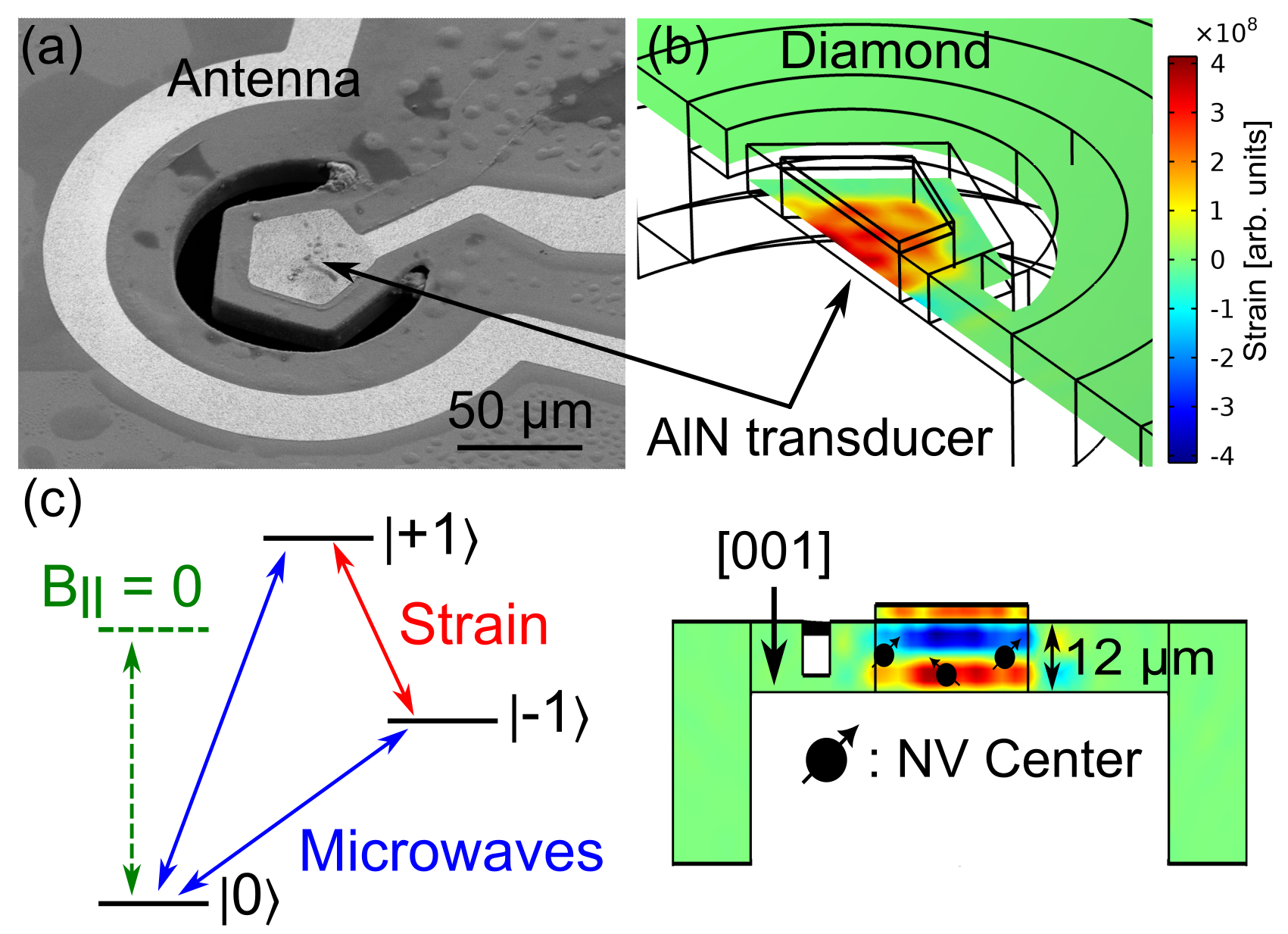}
\caption{{\label{fig:Device}} (a) SEM image of the pentagonal FBAR and MW antenna. (b) Finite-element (COMSOL) depth and cross-sectional models (sharing the same scale) of the pentagonal diamond FBAR qualitatively showing the strain profile. (c) Energy levels of the NV center ground state with a non-zero field along the N-V axis, $B_{||}$, along with corresponding driving fields used in this work.}
\end{figure}

We fabricate thin-film bulk acoustic resonators (FBARs) that are thinner than previous generations of diamond HBAR devices for acoustic NV center control \cite{Macquarrie_2013, Macquarrie_2015, Chen_2019, Chen_2020}. This increases the strain coupling to the NV centers because the coupling is inversely proportional to the square root of the modal volume ($g \propto 1/\sqrt{V}$) \cite{Chen_2019}. The fabrication starts with a 3 mm $\times$ 3 mm $\times$ 50 $\mu$m thick double side polished single crystal diamond with a roughly uniform ensemble of NV centers, purchased from Element Six. The diamond is back-side etched down to 12 $\mu$m in preparation for the FBAR device structures. The 1.5 $\mu$m thick pentagonal AlN transducer layer is sandwiched between two 100 nm platinum electrodes along with 10 nm of titanium for adhesion between the layers. We also fabricate a microwave (MW) antenna on the diamond to provide MW control over the NV centers. Fig.~\ref{fig:Device}(a) shows an SEM image of the final device. 

Exciting the resonator at gigahertz frequencies with an AC voltage excites standing bulk acoustic waves into the diamond with the entire structure thickness acting as an acoustic Fabry-P\'erot cavity \cite{Macquarrie_2013}. These GHz bulk waves are useful for controlling NV center spins in the diamond \cite{Macquarrie_2013, Macquarrie_2015, Chen_2019, Chen_2020}. Fig.~\ref{fig:Device}(b) shows a qualitative finite-element model of a GHz strain wave in depth and a cross-section of the device. There are three strain antinodes present in the 2.553 GHz mode used for all the measurements in this work; one is mainly located in the AlN transducer (dark red color) and the other two in the depth of the diamond (dark blue and red colors). We use this model later to aid in the X-RAY diffraction analysis in Sec.~\ref{sec:XRAY} as a first guess of the strain profile in depth.

The fabricated diamond chip is wire-bonded to printed circuit boards (PCBs) for delivering the necessary microwave and mechanical drive to the device for all measurements [Fig.~\ref{fig:Device}(c)]. Once bonded, we measure the \emph{in situ} scattering parameters (S-parameters) using a vector network analyzer (VNA) to locate the resonance mode frequencies. When driven at a mode frequency, the applied power is transferred to an acoustic excitation of the diamond through the bulk. We use a modified Butterworth Van-Dyke model \cite{Feld_2008,Lason2000} to fit the reflected power ($S_{11}$) VNA data of the 2.553 GHz mechanical mode, giving a quality factor of Q $\approx$ 200. The following sections describe measurements of the strain in the diamond produced by this mechanical mode with both scanning X-ray diffraction microscopy and NV Rabi oscillations in a confocal microscope.

\section{Results \label{sec:Results}}

\subsection{Stroboscopic Scanning X-ray diffraction microscopy}\label{sec:XRAY}

\begin{figure}[ht]
\includegraphics[scale=1]{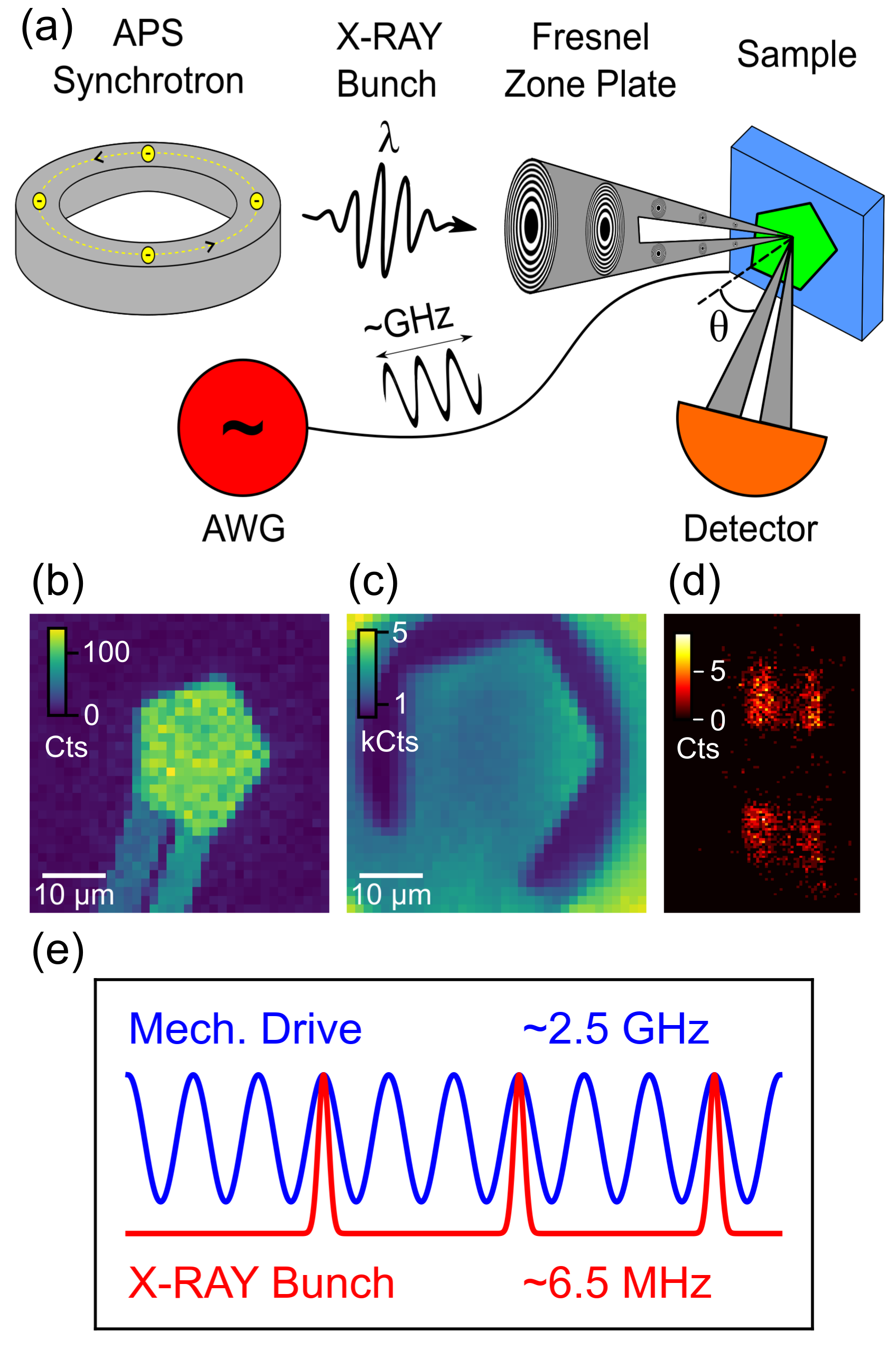}
\caption{{\label{fig:APS}} (a) Schematic of the experimental setup at the Advanced Photon Source. Focused X-rays diffract onto a detector while mechanically driving the FBAR with the AWG. (b) Platinum florescence showing the pentagonal electrode (c) X-ray diffraction image showing the pentagonal FBAR structure, where each pixel is the sum of the corresponding detector fluorescence. (d) Example pixelated area detector image for the center pixel of scan (c). Each pixel of the diffraction scan has a corresponding detector image like the one shown here. (e) Phase synchronous driving of FBAR and X-ray bunch. Mechanical drive is driven at a frequency multiple of the $\sim$6.5 MHz bunch clock.}
\end{figure}

The Advanced Photon Source generates hard X-rays for our experiment in a configuration shown schematically in Fig.~\ref{fig:APS}(a). The 11.6 keV synchrotron pulses are focused down to a $\sim$25 nm FWHM pencil-like, Gaussian profile \cite{Hruszkewycz_2014,Delegan_2023} using a Fresnel zone plate optic \cite{Winarski_2012}. The zone plate is scanned relative to the sample position with 25 nm typical step size in x and y directions \cite{Delegan_2023}, with z positioned along the beam direction. We collect the reflected diffraction from the [113] Bragg condition of diamond in the far field on a two-dimensional Eiger2 X 1M Pixel Array Detector (PAD). We simultaneously measure the platinum fluorescence of the pentagonal electrodes of the FBAR to locate the position on the sample [Fig.~\ref{fig:APS}(b)] by collecting fluorescence on another detector. We simultaneously measure the corresponding X-RAY diffraction images at the same location [Fig.~\ref{fig:APS}(c)] where each pixel is the total sum of diffraction counts on the PAD [Fig.~\ref{fig:APS}(d)]. The detector image appears split in half vertically due to a central stop that is concentric with the focusing optic (this blocks the zeroth order unfocused x-ray beam creating an annular illumination), and split horizontally along the detected Bragg scattering angle (conventional $2\theta$) direction due to a strain variation within the illuminated volume.  The vertical streaks observed on the detector represent an elongation of the diffraction pattern out of the scattering plane due to the focusing optic and are summed in the analysis due to their self-similar two theta values \cite{Hruszkewycz_2014, Hruszkewycz_2015, Pateras_2018}.

The microwave excitation of the resonator is synchronized to a multiple of the X-ray bunch timing of $\sim$6.5 MHz \cite{StorageRing} [Fig.~\ref{fig:APS}(e)] by taking the bunch clock as an external reference clock of the arbitrary waveform generator (AWG), which directly synthesizes the microwave drive tone. This allows for stroboscopic measurements of the mechanical drive via the collected diffraction images. With picosecond temporal control over the relative phase between the X-ray bunches and the mechanical drive, we measure the entire oscillating strain wave at the same physical location on the sample. Because X-rays are weakly interacting with materials \cite{Holt_2013} and our diamond is only 12 $\mu$m thick, our diffraction images are summations of the diffraction through the entire thickness of our diamond. This allows us to directly study the effects of bulk acoustic waves using X-ray diffraction. 

\begin{figure*}
\includegraphics[scale=1]{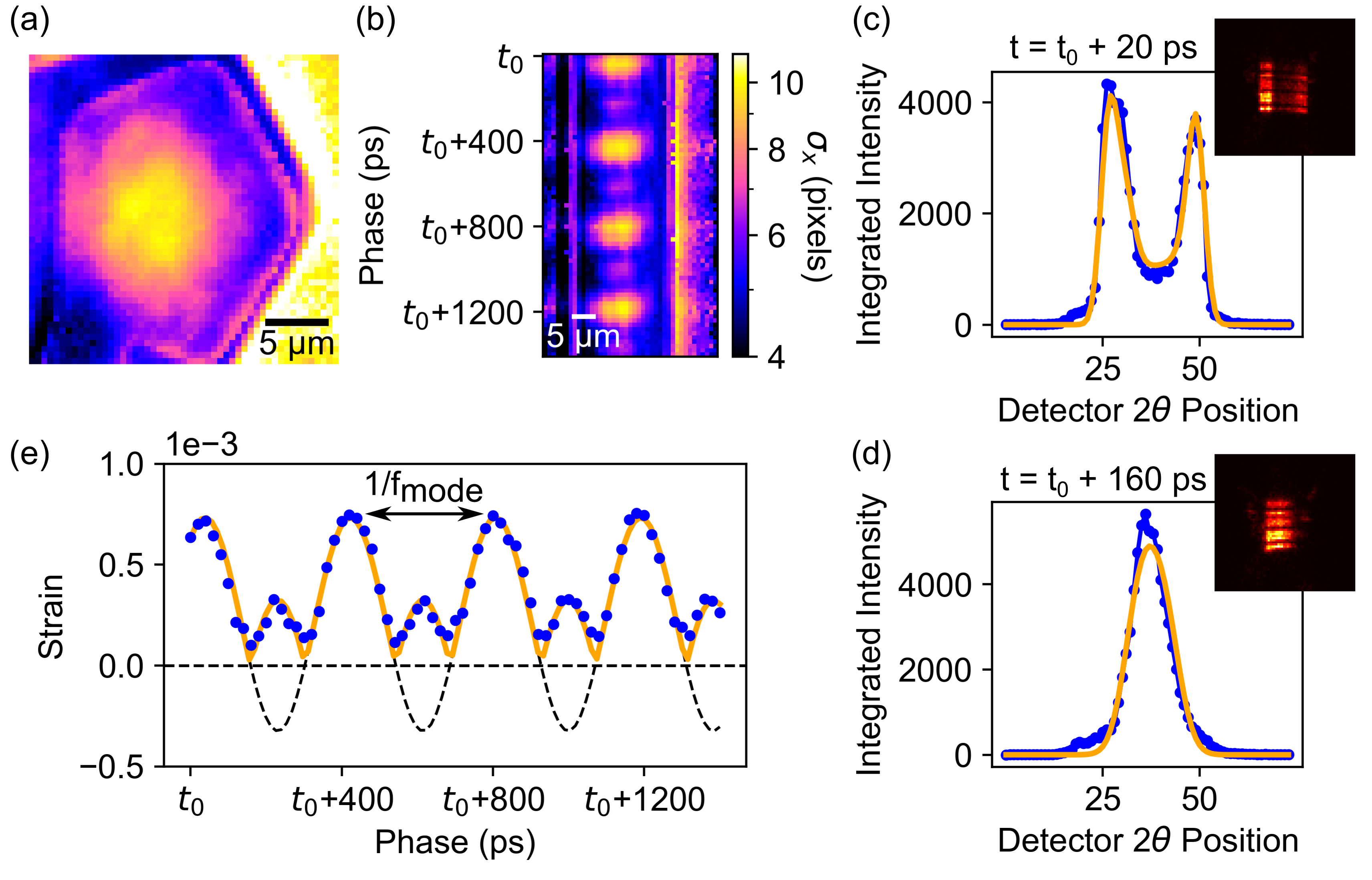}
\caption{{\label{fig:Strain_Fitting}}Results of X-ray diffraction microscopy strain measurements. (a) Two-dimensional X-ray diffraction scan of the diamond FBAR at a fixed phase value showing the horizontal standard deviation $\sigma_x$ of the detector image at each pixel. (b) One-dimensional line-cut across the center of the FBAR versus phase relative to a starting time $t_0$. Both share the same scale. (c) and (d) Strain fits (orange) for the vertically summed data (blue) of each of the corresponding detector images in the top right corner for the phase values of $t = t_0$ + 20 ps and $t = t_0$ + 160 ps, respectively. (e) Strain amplitude fit where each blue dot is the calculated strain amplitude value at the corresponding phase. The fit function is Eq.~\ref{eq:strainfitfunc}, while the dashed line is the function with no absolute value.}
\end{figure*}

The results of the stroboscopic scanning X-ray diffraction measurements with mechanical drive are shown in Fig.~\ref{fig:Strain_Fitting}. The acoustic wave creates both compressive and tensile strain through the bulk that causes lattice compression and expansion. This creates a changing lattice spacing $d$ through the bulk of the diamond. The first-order Bragg condition, $2d\sin(\theta)$ = $\lambda$ \cite{Bragg_1913}, gives the relationship between $d$ and the angle $\theta$ of the measured diffraction, at a wavelength $\lambda$.  Therefore, measuring variations in the Bragg angle gives variations in the lattice spacing under AC strain. Then, we calculate the applied strain using $\epsilon = (d'-d)/d$, the relative difference of the previous and new lattice spacings, $d$ and $d'$ respectively. Since we are summing over diffraction of the entire bulk, we measure diffraction integrated over both the compressive and tensile strain from the bulk acoustic wave at the same time. This creates two lobes on the diffraction patterns that oscillate with the changing phase of the X-ray bunches [Fig.~\ref{fig:Strain_Fitting}(c) and Fig.~\ref{fig:Strain_Fitting}(d)].

To visualize this dynamic strain in images, we use the horizontal standard deviation in the detector images, $\sigma_x$, as a proxy. As the strain amplitude increases, the distance between the compressive and tensile lobes increases, resulting in a larger $\sigma_x$. Fig.~\ref{fig:Strain_Fitting}(a) shows a $15 \times 15$~$\mu m$ scan ($51 \times 51$ pixels) of the FBAR with 0.4~s collection time at each pixel. Then, $\sigma_x$ is calculated at each pixel after summing vertically over the corresponding detector image. The largest region of bulk acoustic strain is about 5 nm wide at the center of the pentagonal FBAR structure. We see very little strain present at the edges of the FBAR and near the anchor, indicating very little loss through the anchor. Fig.~\ref{fig:Strain_Fitting}(b) is a 40 $\mu$m wide 1D plot of diffraction in the x-direction across the center of the FBAR. We repeat this measurement from the starting time $t_0$ to $t = t_0 + 1400$ ps with 20 ps step sizes of the X-ray bunches relative to the mechanical drive with 1 s of collection time at each pixel. We can see the maximal strain region towards the center of the FBAR again. The mechanical drive is performed at a strain frequency of $f_{APS} =$ 2.548 GHz, which is 391 times the X-ray bunch frequency. Therefore, we expect to see multiple oscillations of the strain wave amplitude over 1400 ps of phase since the period is $T = 1/f \approx 400$ ps.

We now present a toy model for fitting the collected diffraction and extracting the strain amplitude from the detector images at each phase. The functional form, $F(x)$, for the fits seen in \ref{fig:Strain_Fitting} (c) and (d) is a convolution of the sinusoidal strain wave in the diamond's thickness with an idealized Gaussian detector response given by
\begin{equation}\label{eq:fitfunc}
\begin{split}
F(x) &= A\int_{0}^{L} G(x-\epsilon(z)) \,dz\,,\\
G(y) &= \frac{1}{\sigma\sqrt{2\pi}}e^\frac{(y-\mu)^2}{2\sigma^2},\\
\epsilon(z) &= \epsilon_0\textrm{sin}(2\pi kz) + mz,
\end{split}
\end{equation}
where $x$ is the pixel position on the detector, $A$ is an overall scaling factor, $L$ is the length of the acoustic strain wave in the diamond depth, and $G(y)$ is a normalized Gaussian with expected value $\mu$ and variance $\sigma^2$.   $\epsilon(z)$ is the spatial strain wave in depth $z$ with an amplitude and spatial frequency of $\epsilon_0$ and $k$ respectively and an additional linear static strain of slope $m$. This convolution results in a function with two peaks of width $\sigma$ and separation $\epsilon_0$ pixels, on each edge corresponding to the highest strain amplitude and the turning points. The difference in amplitude of the two peaks results from the relative amounts of the positive to the negative strain in $\epsilon(z)$. We fix the applied dynamic strain to zero at the diamond/air boundary, but let the amplitude be free at the diamond/AlN boundary by integrating over a variable length of oscillations $L$, since we only qualitatively know the strain profile in depth from the finite-element model [Fig.~\ref{fig:Device}(b)]. We also model the linear static strain that we clearly see as the FBAR is bowed out from fabrication in other diffraction images with $m$. We do not add an additional strain offset to correctly set the zero point of this static strain to be somewhere in the bulk, because adding an offset just shifts the functional form along $x$. We let $\mu$ entirely take care of the $x$ position to avoid covariance in the fit because the size of our detector collection area we choose is already arbitrary. Fig.~\ref{fig:StrainDetector} shows both the strain profile in depth $\epsilon(z)$ and an example detector function $F(x)$. A detailed analysis for choosing the fitting parameters is discussed further in Appendix \ref{App:xray}.

Fig.~\ref{fig:Strain_Fitting}(e) shows the results of the fit strain amplitude at the maximum of strain in the center of the FBAR, for each of the 71 detector images at different phase values. Each data point is the amplitude of the AC strain as a function of phase after converting from the splitting in pixels from our fit, $\epsilon_0$, using the change in $d$ spacing, with a full description in Appendix \ref{App:PtoS}. We then use a change of basis transformation to convert the projection of the strain from the [113] direction to the [001] crystal axis, because the FBAR generates stress along the thickness of the diamond, see Appendix \ref{App:COB}. We fit the data with an absolute value of cosine wave with an offset,
\begin{equation}\label{eq:strainfitfunc}
\epsilon_{[001]}(t) = |\epsilon_{[001]} \cos (2\pi ft + \phi_t) + \epsilon_{off}|
\end{equation}
where $\epsilon_{[001]}$ and $\epsilon_{off}$ are the amplitude and offset respectively and $\phi_t$ is the temporal phase value. We use this function because we see a second peak that occurs since the compressive and tensile lobes overlap as they both oscillate at the mode frequency. The model can account for the difference in heights of the two peaks, but it is clear from all the fits shown in Fig.~\ref{fig:AllFits} that the peaks vary in height from phase to phase. Therefore, we just account for the overall splitting between the two peaks and use the absolute value function [Eq.~\ref{eq:strainfitfunc}]. This effect is not predicted by our model, and could be due to optical effect of the optics used in the X-ray experimental setup [Fig.~\ref{fig:APS}(a)]. We extract an amplitude of the mechanical drive of $\epsilon_{[001]} = 5.2877(9) \times 10^{-4}$ at 0.96 W of driving power applied to the device's port. Also, the strain amplitude oscillates at $f=2.60385(8)$ GHz, while we drive at $f_{mode}=2.553$ GHz.

Fig.~\ref{fig:AllFits} shows the fits for all the detector images using the model described by Eq.~\ref{eq:fitfunc}. The model does well in choosing the location and splitting between the two peaks, and we are confident in the extracted amplitude of strain for all data points. However, our simplified model does not capture all the dynamics that occur while measuring the diffraction. For example, the model results in a poor fit of the lower strain amplitude points, once we fix the peak widths $\sigma$ following the procedure described in Appendix \ref{App:xray}. This may be because we are approaching the dynamical diffraction limit due to the thickness of the diamond \cite{Takahashi_1995}. With an asymmetric broadened functional form for dynamical diffraction, we could correct for things like refraction, shape and width of the peaks, and interference effects, overall resulting in a more accurate fit at the cost of additional fitting parameters. We will discuss these results further with the corresponding strain measurements through mechanically-driven Rabi measurements of the nitrogen vacancy center defects in the diamond FBAR.

\subsection{Strain-driven Rabi oscillations of diamond NV centers}\label{sec:Rabi}

The negatively charged nitrogen vacancy center has a spin triplet ground state. In the absence of external electric fields, its spin Hamiltonian is given by 
\begin{equation}\label{eq:GSHam}
H/h = DS_z^2 + \gamma_e\textbf{B}\cdot\textbf{S} + H_\sigma/h,
\end{equation}
where $h$ is Plank's constant, $D$ = 2.87 GHz is the zero-field splitting of the $\ket{0}$ and $\ket{\pm1}$ electron spin states, $\textbf{S} = (S_x,S_y,S_z)$ is the dimensionless spin-1 operator, $\gamma_e$ = 2.802 MHz/G is the electron gyromagnetic ratio, $\textbf{B} = (B_x,B_y,B_z)$ is an applied magnetic field, and $H_\sigma$ is the spin-stress Hamiltonian \cite{Udvarhelyi_2018}. We fabricate a microwave antenna [Fig.~\ref{fig:Device}(a)] to apply out of plane AC magnetic fields with the form, $B = B_0\cos(\omega t)$. This magnetic field can drive the single quantum (SQ) transitions between $\ket{0} \leftrightarrow \ket{1}$ and $\ket{0} \leftrightarrow \ket{-1}$ when the frequency $\omega$ matches the energy difference between the corresponding spin states. The fabricated FBAR drives the double quantum (DQ) transition between $\ket{-1} \leftrightarrow \ket{1}$ with oscillating strain, which is not accessible magnetically [Fig.~\ref{fig:Device}(c)].


The FBAR generates oscillating uniaxial stress along the [001] crystal direction. With the stress, $\sigma_{ZZ}(t) = |\sigma_{ZZ}|\cos(\omega_{m}t)$, only along this direction, the $H_\sigma$ spin-stress Hamiltonian in the NV basis ($\ket{-1}$, $\ket{0}$, $\ket{1}$), reduces to 
\begin{equation}\label{eq:RStressHam}
H_\sigma = 
\begin{pmatrix}
a_1 & \frac{2}{\sqrt{2}}d & -2b \\
\frac{2}{\sqrt{2}}d & 0 & -\frac{2}{\sqrt{2}}d \\
-2b & -\frac{2}{\sqrt{2}}d & a_1
\end{pmatrix}
\sigma_{ZZ},
\end{equation}
where $a_1$ \cite{Udvarhelyi_2018, Barson_2017, Barfuss_2019}, $b$ \cite{Udvarhelyi_2018, Barson_2017, Barfuss_2019}, and $d$ \cite{Udvarhelyi_2018} are spin-stress coupling parameters. A component of the applied stress wave resonantly couples to the  $\ket{-1} \leftrightarrow \ket{1}$ transition, allowing us to mechanically drive coherent Rabi oscillations on the DQ transition \cite{Macquarrie_2013, Macquarrie_2015, Chen_2019, Chen_2020}.  The Rabi frequency is $\Omega_{m} = 2b|\sigma_{ZZ}|$, corresponding to the magnitude of the $\bra{\pm1}H_\sigma\ket{\mp1}$ matrix elements in Eq.~\ref{eq:RStressHam}, after applying the rotating wave approximation in the rotating frame of the applied mechanical drive field. Stress is related to strain through the elastic stiffness tensor \cite{Kings_1993} such that we write the strain along [001] as $\epsilon_{ZZ} = \sigma_{ZZ}/C_{11}$.  Here, $C_{11} = 1079(5)$ GPa \cite{Mcskimin_2003} is the Young's modulus of diamond. Then, we relate the Rabi frequency to strain using $\Omega_{m} = 2bC_{11}|\epsilon_{ZZ}|$. Therefore, a Rabi measurement enables a quantitative measurement of acoustic strain in the resonator. We average over some inhomogeneity as we measure because of the depth-of-focus in our optical collection, which includes NV centers at a small range of different depths with varying strain amplitudes. This inhomogeneity primarily determines the damping of the Rabi oscillations and does not significantly change the Rabi frequency that we observe. We can then compare these indirect strain measurements to the direct response from the X-ray analysis results in Sec.~\ref{sec:XRAY}.

\begin{figure*}
\includegraphics[scale=1]{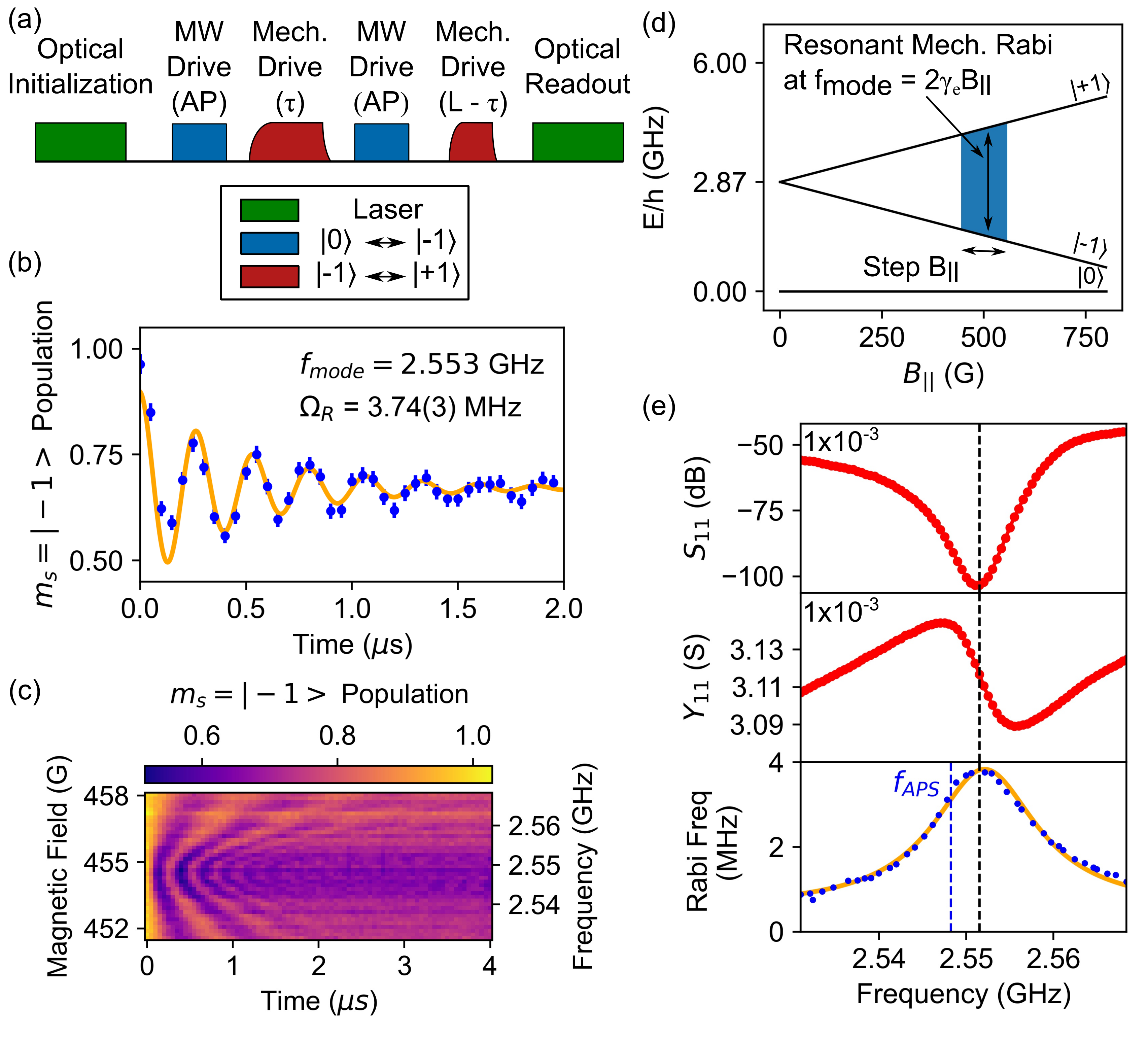}
\caption{{\label{fig:Rabi}}Results of mechanical Rabi measurements. (a) Pulse sequence for mechanical driving \cite{Macquarrie_2015}. (b) Mechanically driven Rabi oscillations between the $\ket{-1}$ and $\ket{+1}$ spin states at 0.96 W of applied mechanical drive power. (c) Resonant mechanical Rabi curves as a function of applied external magnetic field, $B_{||}$, or similarly, the mechanical drive frequency, $f_{mode}$. (d) Ground spin state energies as a function of external field showing the resonant mechanical drive. (e) Comparison of the electrical responses of the reflected power ($S_{11}$) and admittance ($Y_{11}$) to the fit Rabi frequencies of the resonant mechanical drive.}
\end{figure*}

To measure coherent mechanical Rabi oscillations, we take advantage of the optical and spin properties of the NV$^-$ center that allow for optical spin initialization, coherent quantum control between spin states, and optical readout with spin-dependent photoluminescence \cite{Barry_2020}. Fig.~\ref{fig:Rabi}(a) shows the sequence we use for mechanically-driven Rabi oscillations \cite{Macquarrie_2015}. First, we optically initialize the ensemble into spin state $\ket{0}$ with a 532 nm laser. Next, an adiabatic passage (AP) pulse shifts the ensemble to $\ket{-1}$ by adiabatically varying the MW drive frequency through the spin transition frequency to robustly transfer the population into the DQ subspace. Then, we apply the mechanical drive at $f_{mode} = 2\gamma_{e}B_{||}$, the energy difference between the two spin states $\ket{-1}$ and $\ket{+1}$ at the applied external field $B_{||}$, to measure mechanical DQ transitions. We vary the length of the mechanical drive pulse, which changes the percentage of population transferred from one state to the other. A second AP pulse transfers the remaining $\ket{-1}$ population back to $\ket{0}$ for readout. This pulse does not effect the oscillation, since it acts on the DQ subspace after we transferred the relevant spins back to $\ket{0}$. We vary the length of this pulse to ensure equal heating of the sample for each time step. We repeat this 27.5 $\mu s$ sequence many times for each data point of the measurement. Each mechanical Rabi curve is acquired at the same physical location on the sample as the X-ray diffraction analysis, the maximum of strain at the center of the FBAR. The results for a 3.74(3) MHz mechanical Rabi oscillation at the $f_{mode}=2.553$ GHz are shown in Fig.~\ref{fig:Rabi}(b).  The data are fit to a decaying sinusoid to extract the Rabi frequency $\Omega_m$.

We then repeat mechanically-driven Rabi measurements as a function of applied magnetic field $B_{||}$, or resonator drive frequency ($f_{mode}=2\gamma_eB_{||}$), shown in Fig.~\ref{fig:Rabi}(c). At each field $B_\parallel$, we adjust the drive frequency to match the DQ transition frequency, demonstrated graphically in Fig.~\ref{fig:Rabi}(d).  Fig.~\ref{fig:Rabi}(e) shows how the mechanically-driven Rabi frequencies compare to the electrical response measured using a VNA. The electromechanical properties of the mode are well described by a modified Butterworth Van-Dyke (BVD) model \cite{Lason2000}. The first plot is the $S_{11}$ reflected power of the device showing a minimum that corresponds to the mode frequency of the device. The second plot is the admittance calculated from the impedance by $|Y(\omega) = 1/Z(\omega)|$. The admittance contains a maximum and minimum that correspond to the series and parallel resonance frequencies, respectively, of the modified BVD model. The third plot shows the Rabi frequency extracted from a fit of each mechanical Rabi curve.

\begin{table}
  \begin{center}
    \caption{Comparison of our direct measurement of the spin-stress susceptibility parameter $b$ using both the X-ray diffraction and mechanical Rabi analysis to
    the three previous measurements from both density functional theory (DFT) and two different experimental results.}
    \label{tab:results}
    \begin{tabular}{c|c} 

     & \textbf{b [MHz/GPa]} \\
    \hline
    \textbf{This work} & 2.73(2) \\
    \hline
    \textbf{DFT \cite{Udvarhelyi_2018}} & 1.94(2) \\
    \hline
    \textbf{Expt. \cite{Barson_2017}} & -2.3(3) \\
    \hline
    \textbf{Expt. \cite{Barfuss_2019}} & 7.1(8) \\
    \end{tabular}
  \end{center}
\end{table}

We now turn toward measurement of Rabi oscillations as a function of the mechanical drive frequency to properly compare NV-detected dynamic strain to X-ray detected dynamic strain. First we note a subtle point -- the stroboscopic measurements at the APS ensure we cannot drive at the maximum response of the mode, which is roughly at both the minimum of the reflected power and halfway between the maximum and minimum of the admittance [Fig.~\ref{fig:Rabi}(e)]. This is still effective because the mode linewidth is large enough that there is still significant resonant response at the nearest APS harmonic, $f_{APS} = 2.548$ GHz, which is $\sim$4 MHz detuned from the center of the electro-mechanical resonance. We do not have this constraint in Rabi measurements, however, enabling us to measure at finely spaced frequencies. We extract a Rabi frequency of 3.12(2) MHz at $f_{APS}$ from a Lorentzian lineshape fit of the Rabi frequencies at both the same mechanical power used for the APS measurements, 0.96 W, and physical location on the FBAR. With the corresponding measurements of the X-ray dynamic strain $\epsilon_{ZZ}$ and the Rabi frequency $\Omega_{m}$, we can calculate a direct measurement of the spin-stress coupling parameter $b=2.73(2)$ MHz/GPa using $\Omega_{m} = 2bC_{11}|\epsilon_{ZZ}|$. The results of the previous three measurements of this parameter along with our measurement are shown in Table \ref{tab:results}.

Our direct measurement of this spin-stress coupling parameter $b$ is in reasonable agreement with the previous measurements. There is an additional systematic error present in the power difference applied to the FBAR between the two analysis methods in Sec.~\ref{sec:XRAY} and Sec.~\ref{sec:Rabi}. Since the two experimental setups are not compatible with each other, we require different PCBs for each measurement. This results in different wire-bond lengths, which can change the electrical response and power delivery to the device. These two wire-bonds have different inductances, leading to an impedance difference that contributes to additional reflected power at the device. We calculate an upper bound of this power difference to be on the same order of the error in the calculated spin-stress coupling parameter $b=2.73(2)$ MHz/GPa. This power difference along with the fitting discussion at the end of section \ref{sec:XRAY} are the main sources of systematic error in our measurements.

\section{Conclusion}\label{sec:Conc}

We perform measurements of gigahertz frequency acoustic strain in bulk acoustic wave resonators using stroboscopic X-ray diffraction microscopy. This technique provides a direct measurement of a bulk acoustic wave summed over the entire diamond depth, allowing for a quantitative measurement of both the applied acoustic strain amplitude and the static strain present from fabrication. Also, we image the modal structure of the resonator, indicating areas with the largest and smallest strain. In addition to diffraction measurements, we perform optical measurements of the Rabi precession frequency of the ensemble of NV centers present in the bulk of the diamond resonator. Since we perform these measurements at corresponding lateral positions and power, we correlate the results to directly measure the spin-stress coupling parameter $b$. Our measurement is in reasonable agreement with the previous measurements of this parameter. The SXDM technique provides new insights on the structure of the diamond resonators and helps to enhance future fabrication of these devices for improved quantum control of defects in diamond.

\begin{acknowledgments}

We thank Brendan McCullian for helpful insights and discussions. Measurements and data analysis was supported by the DOE Office of Science (National Quantum Information Science Research Centers).  Device development and fabrication was supported by the DARPA DRINQS program (Cooperative Agreement \#D18AC00024).  Device fabrication was performed in part at the Birck Nanotechnology Center at Purdue and at the Cornell Nanofabrication Facility (CNF).  CNF is a member of the National Nanotechnology Coordinated Infrastructure (NNCI), which is supported by the National Science Foundation (Grant No. NNCI-2025233).  Device fabrication was also performed in part at the Cornell Center for Materials Research Shared Facilities that are supported through the NSF MRSEC program (Grant No. DMR-1719875). Work performed at the Center for Nanoscale Materials and Advanced Photon Source, both U.S. Department of Energy Office of Science User Facilities, was supported by the U.S. DOE, Office of Basic Energy Sciences, under Contract No. DE-AC02-06CH11357. The views, opinions and/or findings expressed are those of the authors and should not be interpreted as representing the official views or policies of the Department of Defense or the U.S. Government. Sunil Bhave performed this research at Purdue University prior to becoming a DARPA program manager.

\end{acknowledgments}

\appendix
\section{X-ray Strain Analysis Fitting}\label{App:xray}

The model used for fitting the detector images, Eq.~\ref{eq:fitfunc}, includes two free parameters for the strain profile other than the amplitude $\epsilon_0$: the length of the acoustic strain wave in the diamond depth $L$ and linear slope of static strain $m$. The length determines the number of nodes/antinodes of the strain wave in the diamond thickness, and $m$ determines the slope of the linear static strain from fabrication. In combination, these two parameters describe the entire spatial profile of the strain wave. However, the strain profile in depth should only vary in amplitude as the phase relative to the X-ray bunch varies. Therefore, to correctly fit the detector images, we have to constrain both of these parameters before fitting each phase value.

We fit five of the highest strained detector images per oscillation with all six fitting parameters free, and take the mean giving $L$ = 0.907(5) oscillations and $m$ = 24(1) pixels. Both of these calculated values are reasonable. We constrain to these values moving forward. $L$ = 0.907(5) matches the qualitative finite-element model of about one full oscillation ($L$ = 1) of the strain amplitude in the diamond [Fig.~\ref{fig:Device}(b)]. Also, we can convert the linear static strain of $m$ = 24(1) pixels to strain using the method in Appendix \ref{App:PtoS}, resulting in a strain of $\epsilon_{off} \approx 2$ x $10^{-3}$. Additionally, with Euler-Bernoulli beam theory, we can approximate the static strain from fabrication. The X-ray diffraction maximum varies over the 50 $\mu m$ long FBAR by $\sim$ 0.4$^\circ$ as we rotate the angle of the sample relative to the X-ray beam. We calculate the radius of curvature of the statically strained FBAR, and using the thickness of the diamond, we arrive at a similar static strain of $\epsilon_{off} \approx 2$ x $10^{-3}$. This agreement supports that the model accurately captures the length of the spatial strain wave as well as the static strain through the depth.

\begin{figure}
\includegraphics[scale=1]{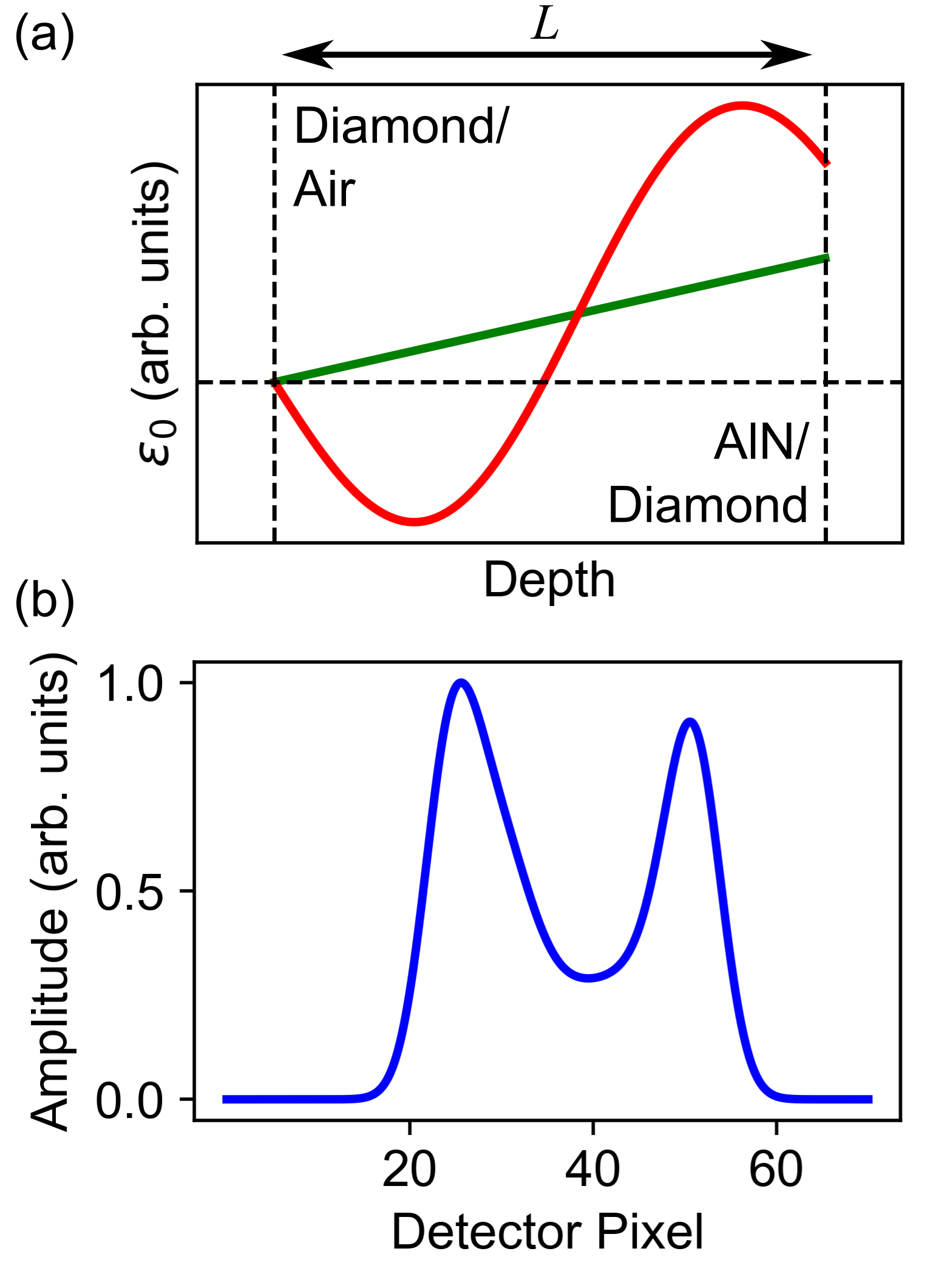}
\caption{{\label{fig:StrainDetector}} Plots the functions used to fit the detector images. (a) The spatial strain profile as a function of depth in the diamond with the constrained period $L$ and linear static strain $m$. (b) Model from Eq.~\ref{eq:fitfunc} with fixed $L$ and $m$ values from our constrains with arbitrary other fitting parameters $\mu, \sigma, \epsilon_0$ and $A$.}
\end{figure}

Fig.~\ref{fig:StrainDetector}(a) shows the sinusoidal spatial strain profile through the depth of the diamond with the constrained length $L$, in addition to the the static strain $m$ term of the model. In practice, the strain would be zero somewhere in the bulk, not at the diamond/air boundary as shown in Fig.~\ref{fig:StrainDetector}(a). However, adding an additional static strain offset term to correct the strain profile just results in a translation of the fit function $F(x)$ along $x$, not changing the shape of the fit function. We let $\mu$ be the sole determination of the $x$ position to avoid covariance between the offset strain and $\mu$. Fig.~\ref{fig:StrainDetector}(b) shows the detector fit model $F(x)$ after constraining $L$ and $m$ with arbitrary $\mu, \sigma, \epsilon_0$ and $A$ values for demonstration.

\begin{figure*}
\includegraphics[scale=1]{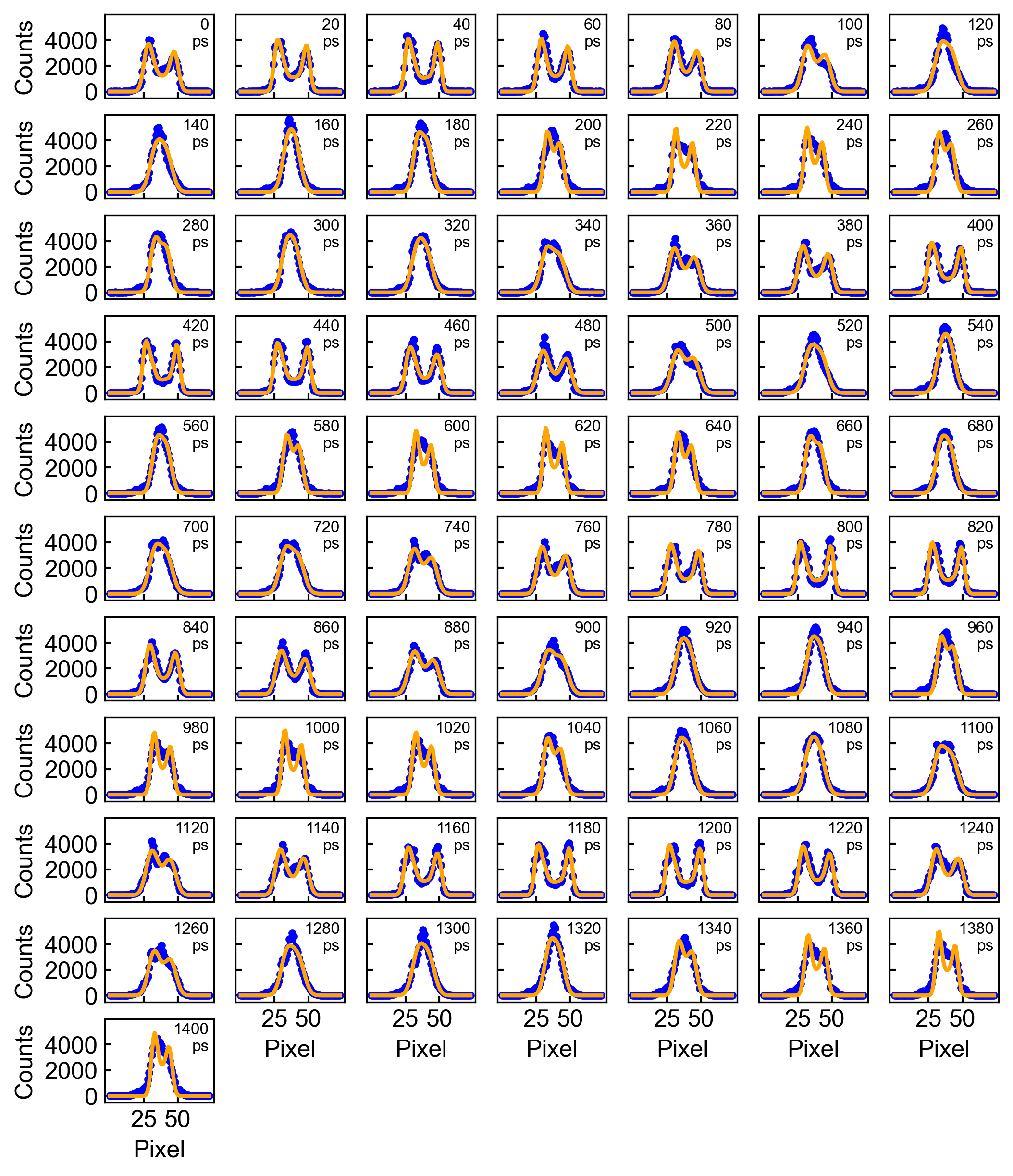}
\caption{{\label{fig:AllFits}} This is a fit of all seventy-one detector images over the varying phase values using the model given in Eq.~\ref{eq:fitfunc} while constraining the parameters $\sigma$, $L$, and $m$ as described by Appendix \ref{App:xray}. The phase value in picoseconds is the time between the start of the mechanical drive and the incoming X-ray bunch.}
\end{figure*}

Additionally, the detector response width $\sigma$ needs to be constrained. There is a large covariance between the width $\sigma$ and the amplitude $\epsilon_0$ for small values of splitting when the two detector lobes are close to each other because a single wide Gaussian can also fit despite not being physical. Therefore, we have to determine $\sigma$ for each phase value of the drive. The $\sigma$'s vary with phase because the temporal width ($\Delta_t$) of the X-ray bunch timings result in a broadened detector response. We define a strain velocity $\epsilon_v(t)$ that corresponds to how much the strain wave is changing in time at each point in the depth, which is maximized at the antinodes and minimized at the nodes. The strain velocity is the derivative of the AC part of the spatial strain wave $\epsilon(z)$ from Eq.~\ref{eq:fitfunc},

\begin{equation}\label{eq:strainvel}
\begin{split}
\epsilon_v(t) &= \frac{d\epsilon(z)}{dt} \\
&= \frac{d\epsilon_0(t)}{dt}\sin(2\pi kz) \\
&= -2\pi\epsilon_{001} f\sin(2\pi ft+\phi_t)\sin(2\pi kz),
\end{split}
\end{equation}
where the amplitude of the spatial strain wave $\epsilon_0(t)$ is time-varying according to Eq.~\ref{eq:strainfitfunc}. At the nodes, there is no broadening from the temporal width of the X-ray bunches. Also, we know the strain velocity is maximized at the antinodes, resulting in $|\epsilon_v| = 2\pi\epsilon_{001} f$. To calculate the broadening of the detector function in pixels, we use $\sigma_{broad} = |\epsilon_v| \Delta_t$, where the $\Delta_t$ = 33.5(1) ps \cite{StorageRing}. Therefore, at the antinodes, we will have the most broadening, and the sigma at those phase values will be given by

\begin{equation}\label{eq:maxsigma}
\sigma_{antinode}^2 = \sigma_{node}^2 + \sigma_{broad}^2.
\end{equation}
Then, $\sigma(t)$ oscillates between $\sigma_{antinode}$ and $\sigma_{node}$ as a cosine wave with the same frequency and phase as Eq.~\ref{eq:strainfitfunc}, giving
\begin{equation}\label{eq:sigmafunc}
\begin{split}
\sigma(t) &= \sigma_0 \cos(2\pi ft+\phi_t) + \sigma_1,\\
\sigma_0 &= \frac{\sigma_{antinode}-\sigma_{node}}{2},\\
\sigma_1 &= \frac{\sigma_{antinode}+\sigma_{node}}{2},
\end{split}
\end{equation}
where $t$ is the time corresponding to a phase between the X-ray bunch and mechanical drive.

With the fixed parameters including the widths and spatial strain parameters (length and linear static strain), the seventy-one detector images with varying phase are fit according to the model described by Eq.~\ref{eq:fitfunc}, and the fits are shown in Fig.~\ref{fig:AllFits}. The strain amplitude $\epsilon_0$ is extracted from each of these fits and plotted as shown in Fig.~\ref{fig:Strain_Fitting}(e).

\section{Converting pixel position on detector to strain}\label{App:PtoS}

The bulk acoustic waves from the FBAR change the lattice throughout the depth of the diamond. We can relate the position of the compressive and tensile lobes on the detector from the X-ray Bragg diffraction to a change in Bragg angle. We know the angle relates to the unstrained lattice spacing $d$ through the first-order Bragg condition~\cite{Bragg_1913},
\begin{equation}\label{eq:BraggApp}
2d\sin(\theta_B) = \lambda, 
\end{equation}
where $d$ = 1.075 {\AA} is the [113] lattice spacing using the diamond lattice parameter $a$ = 3.567 {\AA} \cite{Prelas_1995}, $\lambda$ = 1.0688(5) {\AA} (11.600(5) keV) is the wavelength of the X-rays, and $\theta_B = 29.81(2) ^{\circ}$ is the Bragg angle. While driving the FBAR, the positions of the two lobes change, and a new Bragg condition results from a new lattice spacing:
\begin{equation}\label{eq:BraggNewApp}
2d'\sin(\theta_B') = \lambda,
\end{equation}
where $d'$ and $\theta_B'$ are the lattice spacing and Bragg angle under mechanical strain respectively. We measure the strain generated by the FBAR along the [113] crystal direction by calculating the relative difference of the unstrained and strained lattice spacings using
\begin{equation}\label{eq:dstrainApp}
\epsilon_{[113]} = \frac{d'-d}{d}.
\end{equation}

To extract the strain from the diffraction images, we calculate $\theta_p$, the change in Bragg angle per pixel, where the pixel corresponds to the location of the diffraction lobes on the detector. We calculate the sample-detector distance by performing a measurement where we rotate the sample by 0.9$^\circ$ with a stage of $\pm\ 24\ \mu$rad angular motion repeatability and measure the diffraction pattern shifting a lateral distance of 234 pixels with a 75 $\mu m$ pixel spacing. Using the law of cosines on an acute isosceles triangle, where two sides are the sample-detector distance and the third is $234\times75\ \mu m$ with an opposite angle of 0.9$^\circ$, the sample-detector distance is 1.1(1) m. Then, forming an acute right triangle with one side as the sample-detector distance and the other as the pixel-pixel distance, we calculate $\theta_p$ = tan$^{-1}$(75 $\mu$m/1.1 m) = 0.0038(3)$^\circ$. Now, we can write the new Bragg condition angle,
\begin{equation}\label{eq:thetaprimeApp}
\theta_B' = \theta_B + \theta_p(\epsilon_0/2),
\end{equation}
where $\epsilon_0$ is the number of pixels between the two peaks from our model for the fits in Fig.~\ref{fig:AllFits}. We include the factor of two since the change in Bragg angle corresponds to the angle from the original Bragg angle to one of the detector lobes, and $\epsilon_0$ is the total distance between both lobes.

Therefore, the fitting of the X-ray diffraction images allow us to calculate the strained Bragg condition angle $\theta_B'$ using Eq.~\ref{eq:thetaprimeApp}, and then the lattice spacing under strain $d'$ using Eq.~\ref{eq:BraggNewApp}. Then, we calculate $\epsilon_{[113]}$, the strain along the [113] crystal direction, through the relative difference of the strained and unstrained lattice spacings, Eq.~\ref{eq:dstrainApp}.

\section{Change of Basis}\label{App:COB}

The FBAR creates uniaxial stress $\sigma$ along the [001] crystal direction via the bulk acoustic waves. With crystal frame coordinates of $X=[100],\ Y=[010],$ and $Z=[001]$, we can write the stress generated from the FBAR as

\begin{equation}\label{eq:FBARstress}
\sigma_{FBAR} = \begin{pmatrix}
0 & 0 & 0\\
0 & 0 & 0\\
0 & 0 & \sigma_{ZZ}
\end{pmatrix},
\end{equation}
\newline
in the crystal frame's coordinate system. The strain $\epsilon$ is related to the stress through the elastic stiffness tensor \cite{Kings_1993}, and we can write the strain generated by the FBAR in the crystal frame as

\begin{equation}\label{eq:FBARstrain}
\begin{split}
\epsilon_{FBAR} &= \begin{pmatrix}
\epsilon_{XX} & 0 & 0\\
0 & \epsilon_{YY} & 0\\
0 & 0 & \epsilon_{ZZ}
\end{pmatrix}\\
 &= \begin{pmatrix}
-\nu & 0 & 0\\
0 & -\nu & 0\\
0 & 0 & 1
\end{pmatrix}\epsilon_{ZZ}\\
 &= \begin{pmatrix}
-\nu & 0 & 0\\
0 & -\nu & 0\\
0 & 0 & 1
\end{pmatrix}\sigma_{ZZ}/C_{11},
\end{split}
\end{equation}
where $|\nu|=C_{12}/(C_{11}+C_{12}) = 0.103(2)$ is Poisson's ratio and $C_{11}=1079(5)$ GPa and $C_{12} = 124(5)$ GPa are elastic stiffness moduli for diamond \cite{Mcskimin_2003}.

We choose new normalized coordinates with the new z direction along [113] crystal direction matching the incoming X-ray angle: $z=[113]/\sqrt{11}$, $x=[1\overline{1}0]/\sqrt{2}$, $y= \overrightarrow{z}$ x $\overrightarrow{x} = [33\overline{2}]$. Using a change of basis transformation \cite{Aadnoy_2019}, 

\begin{equation}\label{eq:COB}
\epsilon' = Q\epsilon Q', \sigma' = Q\sigma Q',
\end{equation}
where $Q$ is a unitary transformation between the two frames,
\begin{equation}\label{eq:Qmatrix}
Q = \begin{pmatrix}
x\\y\\z
\end{pmatrix}
\begin{pmatrix}
X\\Y\\Z
\end{pmatrix}^T,
\end{equation}
the applied strain from the FBAR in the new basis will be given by
\begin{equation}\label{eq:StrainNV}
\epsilon' = \begin{pmatrix}
-\nu & 0 & 0\\
0 & \frac{-9\nu+2}{11} & \frac{-3\sqrt{2}(\nu+1)}{11}\\
0 & \frac{-3\sqrt{2}(\nu+1)}{11} & \frac{-2\nu+9}{11}
\end{pmatrix}\sigma_{ZZ}/C_{11}.
\end{equation}
Therefore, we can convert the strain the [113] crystal direction to the [001] direction using the corresponding component of $\epsilon'$ giving 

\begin{equation}\label{eq:FBARstrain001}
\begin{split}
\epsilon_{[113]} &= [(-2\nu+9)/11]\sigma_{ZZ}/C_{11}\\
&= [(-2\nu+9)/11]\epsilon_{ZZ}\\
&= [(-2\nu+9)/11]\epsilon_{[001]}.
\end{split}
\end{equation}

\nocite{*}

\bibliography{bibliography}

\end{document}